\documentclass[letterpaper, 10 pt, conference]{ieeeconf}  %

\IEEEoverridecommandlockouts                              %

\usepackage{graphicx} %
\usepackage{times} %
\usepackage{amsmath} %
\usepackage{amssymb}  %
\usepackage{amsfonts}

\usepackage{mathtools}
\usepackage[hidelinks]{hyperref}
\usepackage{amsthm}
\usepackage{enumerate}
\usepackage[ruled, linesnumbered]{algorithm2e}
\usepackage{xcolor}
\usepackage{multicol, multirow}
\usepackage{setspace}
\newtheorem{remark}{Remark}
\usepackage{tikz}
\newcommand*\textcircle[1]{\tikz[baseline=(char.base)]{
            \node[shape=circle,draw,inner sep=0.9pt] (char) {#1};}}

\newcommand{\WP}[2]{\left\llbracket{#1}, {#2}\right\rrbracket}

\DeclareSymbolFont{bbold}{U}{bbold}{m}{n}
\DeclareSymbolFontAlphabet{\mathbbold}{bbold}

\newcommand{\real}{\mathbb{R}}

\usepackage{stmaryrd}

\newcommand{\seminorm}[1]{{\left\vert\kern-0.25ex\left\vert\kern-0.25ex\left\vert #1
		\right\vert\kern-0.25ex\right\vert\kern-0.25ex\right\vert}}

\newcommand{\semimeasure}[1]{\mu_{\seminorm{\cdot}}\kern-0.5ex\left(#1\right)}

\DeclareMathOperator{\diag}{diag}

\renewcommand{\top}{\mathsf{T}} %

\title{\LARGE Contraction-Guided Adaptive Partitioning for Reachability Analysis of Neural Network Controlled Systems }

\author{Akash Harapanahalli, Saber Jafarpour, and Samuel Coogan%
\thanks{*This work is supported in part by Cisco Systems, the National Science Foundation under Grants \#1749357 and \#2219755, the Air Force Office of Scientific Research under Grant FA9550-23-1-0303, and  NASA under the University Leadership (ULI) Award \#80NSSC20M0161 but solely reflects the opinions and conclusions of its authors.}
\thanks{Akash Harapanahalli, Saber Jafarpour, and Samuel Coogan are with the School of Electrical and Computer Engineering, Georgia Institute of Technology, USA, {\tt\small \{aharapan,saber,sam.coogan\}@gatech.edu}}
}

\newcommand{\R}{\mathbb{R}}
\newcommand{\set}[1]{\mathcal{#1}}
\newcommand{\h}[1]{\widehat{#1}}

\definecolor{tab:blue}{HTML}{1f77b4}
\definecolor{tab:orange}{HTML}{ff7f0e}
\definecolor{tab:green}{HTML}{2ca02c}
\definecolor{tab:red}{HTML}{d62728}
\definecolor{tab:purple}{HTML}{9467bd}
\definecolor{tab:brown}{HTML}{8c564b}
\definecolor{tab:pink}{HTML}{e377c2}
\definecolor{tab:gray}{HTML}{7f7f7f}
\definecolor{tab:olive}{HTML}{bcbd22}
\definecolor{tab:cyan}{HTML}{17becf}

\newtheorem{theorem}{Theorem}
\newtheorem{assumption}{Assumption}

\SetKwInOut{Parameter}{Parameter}

\begin{document}
\RestyleAlgo{ruled}

\maketitle
\thispagestyle{empty}
\pagestyle{empty}

\begin{abstract}

In this paper, we present a contraction-guided adaptive partitioning algorithm for improving interval-valued robust reachable set estimates in a nonlinear feedback loop with a neural network controller and disturbances. 
Based on an estimate of the contraction rate of over-approximated intervals, the algorithm chooses when and where to partition. Then, by leveraging a decoupling of the neural network verification step and reachability partitioning layers, the algorithm can provide accuracy improvements for little computational cost. 
This approach is applicable with any sufficiently accurate open-loop interval-valued reachability estimation technique and any method for bounding the input-output behavior of a neural network. 
Using contraction-based robustness analysis, we provide guarantees of the algorithm's performance with mixed monotone reachability.
Finally, we demonstrate the algorithm's performance through several numerical simulations and compare it with existing methods in the literature. In particular, we report a sizable improvement in the accuracy of reachable set estimation in a fraction of the runtime as compared to state-of-the-art methods.

\end{abstract}

\section{Introduction}

Neural networks have become increasingly popular in control systems in recent years due to their relative ease for in-the-loop computation. 
These learning-based algorithms are known to be vulnerable to input perturbations---small (possibly adversarial) changes in their input can lead to large variations in their output~\cite{CZ-WZ-IS-JB-DE-IG-RF:13}.
As such, runtime verification of the safety and performance of neural network controlled systems is essential in safety-critical applications.
This task is generally challenging due to the nonlinear and large-scale structure of the neural networks and their interconnection with nonlinear dynamics~\cite{SD-XC-SS:19}. 

A basic ingredient for verifying control systems is the ability to overapproximate the set of reachable states from a given set of initial conditions, possibly in the presence of disturbances. If, for example, this overapproximation avoids obstacles or reaches a goal region, then the system certifiably satisfies the corresponding safety or performance criteria. 

Recently, several promising reachability-based methods have been proposed for verifying 
stand-alone neural networks or feedback systems with neural networks in the control loop; however, these methods either suffer from large computational complexity, large over-approximation error, or lack of generality.
For stand-alone neural networks: Interval Bound Propagation (IBP)~\cite{SG-etal:19} is fast, but largely over-conservative; CROWN~\cite{HZ-etal:18} propagates linear bounds through the network; LipSDP~\cite{MF-MM-GJP:22} uses semi-definite programming, but is not scalable to large networks; NNV~\cite{HDT-etal:20} is a set-based verification framework implemented in MATLAB. 
For neural network closed-loop verification: ReachLP \cite{ME-GH-CS-JPH:21} is computationally light, but it can be conservative and only applies to linear systems; ReachSDP~\cite{HH-MF-MM-GJP:20} can provide tighter bounds, but does not scale well to large networks and only applies to linear systems; JuliaReach~\cite{CS-MF-SG:22}, CORA~\cite{NK-CS-MA-SB:23}, NNV~\cite{HDT-etal:20}, POLAR~\cite{CH-JF-XC-WL-QZ:22}, MILP~\cite{CS-AM-AI-MJK:22}, and constrained Zonotope~\cite{YZ-XX:22} methods can provide very accurate estimates for non-linear systems, but are too expensive for runtime verification. ReachMM~\cite{SJ-AH-SC:23} combines off the shelf open-loop verifiers with mixed monotonicity to provide efficient but coarse interval over-estimates.

\begin{figure}
    \label{fig:step}
    \centering
    \includegraphics[width=1\linewidth,trim={0.5cm 0 0.5cm 0},clip]{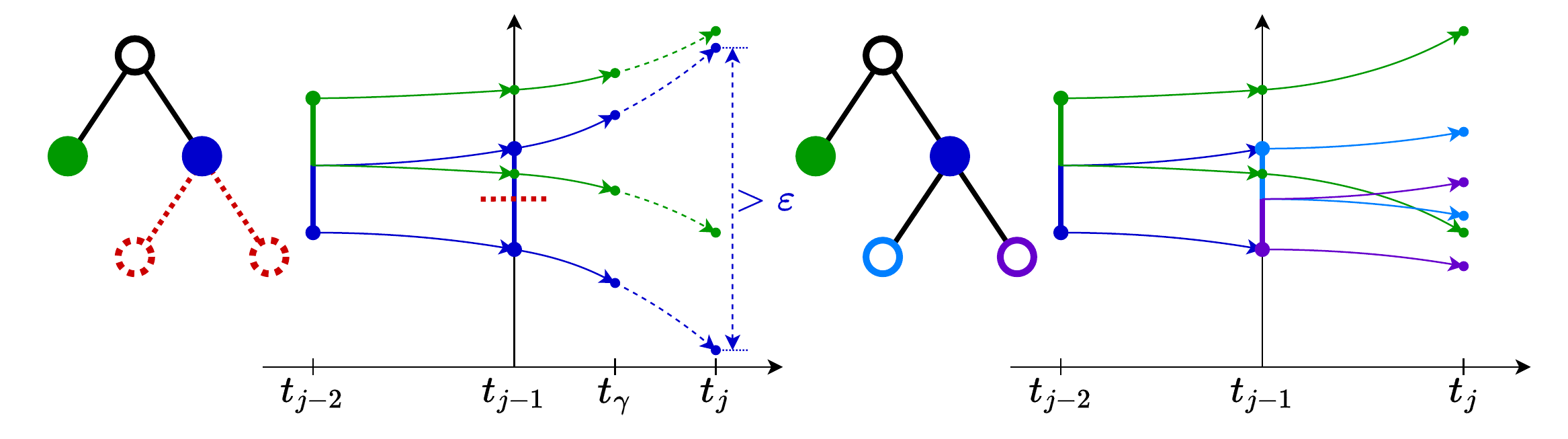}
    \caption{A snapshot of Algorithm~\ref{alg:main} is illustrated for $n=1$. \textbf{(top)} There are two initial partitions that both compute the neural network verification step (filled circles in the graph representation). Both partitions are integrated from $t_{j-1}$ to $t_\gamma=t_{j-1} + \gamma(t_j - t_{j-1})$ to estimate the width of the box at $t_j$. Since the estimated blue partition width violates the user-defined maximum width $\varepsilon$, \textbf{(bottom)} the algorithm returns to $t_{j-1}$ and adds two new partitions to the tree. The maximum neural network verification depth here is $1$, so the new partitions do not recompute the neural network verification and instead use the same $E^c$ \eqref{eq:clembsys} from the initial blue partition.}
    \vspace{-1em}
\end{figure}
Partitioning of the state space is an effective approach for balancing the trade-off between the accuracy of reachable set estimates and the runtime of the verification strategy. 
In the literature, there are many existing partitioning algorithms.
For standalone ReLU networks,~\cite{VRR-RC-DMS-CT:21} splits the input set into partitions based on the stability of each neuron. In~\cite{ME-GH-JPH:21,WX-HDT-XY-TTJ:21}, Monte Carlo simulations are used to guide which specific axes to cut. For closed-loop linear systems, ReachLP-Uniform~\cite{ME-GH-CS-JPH:21} uses a uniform partitioning of the state space, ReachLP-GSG~\cite{ME-GH-CS-JPH:21} applies the GSG algorithm from~\cite{WX-HDT-XY-TTJ:21} to ReachLP, and ReachLipBnB~\cite{TE-SS-MF:22} combines a branch and bound algorithm with $\ell_2$-Lipschitz constants obtained from LipSDP.

\paragraph*{Contributions}

We introduce a contraction-guided adaptive partitioning algorithm for general nonlinear systems with neural network controllers. 
In Section~\ref{sec:theory}, we use contraction theory to prove that for the mixed monotone framework ReachMM~\cite{SJ-AH-SC:23}, the expansion rate of interval partitions is directly linked with the contraction rate of the embedding system. In particular, we provide upper bounds on the width of interval reachable set over-estimates---showing that their exponential growth in time accumulates from three main sources: (1) the maximum rate of expansion of the closed loop system in the specific partition, (2) the initial width of the partition, and (3) the approximation error of the neural network verifier.
This motivates the creation of Algorithm~\ref{alg:main} in Section~\ref{sec:alg}, which provides three key features for improving these sources of error.
The algorithm is 
\textit{separative}---it decouples the neural network verification calls and reachability partitioning layers, leading to inexpensive accuracy benefits on (2) without suffering additional computational cost on improving (3); 
\textit{spatially aware}---it chooses specific regions of the state space to cut based on the contraction rate of the partitions, allocating resources according to where in space (1) is the worst; and 
\textit{temporally aware}---it chooses to cut along trajectories as estimates blow up, allocating resources according to when in time (1) is the worst.
By applying the algorithm to the mixed monotone framework~\cite{SJ-AH-SC:23}, ReachMM-CG is applicable in continuous or discrete time, for nonlinear plant models controlled by neural networks with general nonlinear activation functions; this removes several limitations found in most of the above-cited approaches. 
We find that ReachMM-CG yields reachable set estimates with a 33\% accuracy improvement in a quarter of the time compared to state-of-the-art partitioning algorithms on a benchmark example.
Finally, while our bounds apply for specifically the strategy proposed in~\cite{SJ-AH-SC:23}, we note that the partitioning algorithm itself can also be coupled with any interval-based verification methods, including many of those discussed above, as well as any future advancements. 

\section{Notation and Mathematical Preliminary}

We denote the set of extended real numbers by $\overline{\R} = \R \cup \{-\infty,\infty\}$.
We define the partial order $\leq$ on $\R^n$ as $x\leq y \iff x_i \leq y_i \,\forall i\in\{1,\ldots,n\}$. 
For every $x\leq y$, we define the interval $[x,y]=\{z : x\leq z \leq y\}$. 
For $x,\h{x}\in\R^n$, let $(x,\h{x})\in\R^{2n}$ be their concatenation. 
The southeast partial order $\leq_{\text{SE}}$ on $\R^{2n}$ is induced by $\leq$ on $\R^n$ as follows: $(x,\h{x}) \leq_{\text{SE}} (y, \h{y}) \iff x \leq y \text{ and } \h{y} \leq \h{x}$. 
Define the following: $\set{T}^{2n}_{\geq0} = \{(x,\h{x})\in\R^{2n} : x \leq \h{x}\},\ \set{T}^{2n}_{\leq0} = \{(x,\h{x})\in\R^{2n} : \h{x} \leq x\},\ \set{T}^{2n} = \set{T}^{2n}_{\geq0} \cup \set{T}^{2n}_{\leq0}$. 
We define the vector $x_{[i:\hat{x}]}\in\R^n$ as $\left(x_{[i:\hat{x}]}\right)_j = \begin{cases} x_j, & j\neq i \\ \hat{x}_j & j = i \end{cases}$. 
Define the weighted $\ell_{\infty}$-norm $\|x\|_{\infty,\varepsilon} = \|\diag(\varepsilon)^{-1} x\|_\infty$. In particular, the weighted maximum width of an interval $[\underline{x},\overline{x}]$ is $\|\overline{x} - \underline{x}\|_{\infty,\varepsilon}$. Given a matrix $A\in \real^{n\times n}$, the $\ell_{\infty}$-matrix measure of $A$ is defined by $\mu_{\infty}(A) =\max_{i \in \{1,\ldots,n\}} \{A_{ii} + \sum_{j \neq i} |A_{ij}|\}$. From~\cite[Table~III]{AD-SJ-FB:20o}, we define the weak pairing $\WP{\cdot}{\cdot}_{\infty}:\real^n\times\real^n\to \real$ associated to the norm $\|\cdot\|_{\infty}$ as follows:
\begin{align*}
    \WP{x}{y}_{\infty}= \max_{i \in I_{\infty}(y)} y_ix_i,
\end{align*}
where $I_{\infty}(x) = \{i\in\{1,\ldots,n\}\mid |x_i|=\|x\|_{\infty}\}$.
Consider the following dynamical system 
\begin{align}\label{eq:dynamics}
    \dot{x}=f(x,w)
\end{align}
with state vector $x\in \real^n$ and disturbance vector $w\in \mathcal{W}\subset \real^q$. Given a piecewise continuous curve $w(\cdot)$ where ${w}:[t_0,t]\to \mathcal{W}$, the trajectory of the system starting from $x_0$ at time $t_0$ is given $t\rightarrow\phi_f(t,t_0,x_0,{w}(\cdot))$. Given an initial set $\set{X}$, we denote the reachable set of $f$ at some $t\geq t_0$:
\begin{gather}
    \set{R}_{f}(t,t_0,\set{X},\set{W}) = \left\{
    \begin{aligned}
    \phi&_{f}(t, t_0, x_0, w(\cdot)),\,\forall x_0\in\set{X},\\&w:\R\rightarrow\set{W} \text{ piecewise cont.}
    \end{aligned}
    \right\}
\end{gather}

The dynamical system~\eqref{eq:dynamics} is mixed monotone with respect to the decomposition function $d:\set{T}^{2n}\times\set{T}^{2q}\rightarrow\R^n$ if, for every $i\in \{1,\ldots,n\}$,
\begin{enumerate}[i.]
    \item $d_i(x,x,w,w) = f_i(x,w)$, for every $x\in\R^n,\,w\in\R^q$;
    \item $d_i(x,\h{x},w,\h{w})\leq d_i(y,\h{y},w,\h{w})$, for every $x\leq y$ s.t. $x_i=y_i$, and every $\h{y}\leq \h{x}$;
    \item $d_i(x,\h{x},w,\h{w})\leq d_i(x,\h{x},v,\h{v})$, for every $w\le v$ and every $\h{v}\le \h{w}$.
\end{enumerate}
With a valid decomposition function $d$, one can construct an embedding system associated to~\eqref{eq:dynamics} as follows:
\begin{gather} \label{eq:olembsys}
    \frac{d}{dt}\begin{bmatrix}\underline{x} \\ \overline{x}\end{bmatrix} = 
    \begin{bmatrix} d(\underline{x},\overline{x},\underline{w},\overline{w}) \\ d(\overline{x},\underline{x},\overline{w},\underline{w}) \\ \end{bmatrix} := E(\underline{x},\overline{x},\underline{w},\overline{w}).
\end{gather}

\section{Problem Statement}

Consider a nonlinear continuous-time dynamical system of the form
\begin{gather} \label{eq:nlsys}
    \dot{x}(t) = f(x(t),u(t),w(t)),
\end{gather}
where $x\in\R^{n}$ is the state of the system, $u\in\R^{p}$ is the control input to the system, $w\in\R^{q}$ is a disturbance input to the system, and $f:\R^n\times\R^p\times\R^q \rightarrow \R^{n}$ is a parameterized vector field. We assume that the feedback control policy for the system~\eqref{eq:nlsys} is given by a $k$-layer fully connected feed-forward neural network $N:\R^n\rightarrow\R^p$ as follows:
\begin{gather}\label{eq:NN}
    \begin{gathered}
        \xi^{(i)} = \sigma^{(i-1)} \left(W^{(i-1)}\xi^{(i-1)} + b^{(i-1)}\right),\, i=1,\ldots k \\
        \xi^{(0)} = x,\quad N(x) = W^{(k)} \xi^{(k)} + b^{(k)}
    \end{gathered}
\end{gather}
where $m_i$ is the number of neurons in the $i$-th layer, $W^{(i)} \in \R^{m_i\times m_{i-1}}$ is the weight matrix on the $i$-th layer, $b^{(i)}\in\R^{m_i}$ is the bias vector on the $i$-th layer, $\xi^{(i)} \in \R^{m_i}$ is the $i$-th layer hidden variable and $\sigma_i$ is the activation function for the $i$-th layer. In practice, $N(x(t))$ cannot be evaluated at each instance of time $t$, and, instead, the control must be implemented via, \emph{e.g.}, a zero-order hold strategy between sampling instances. We assume that there exists an increasing sequence of control time instances $\{t_0,t_1,t_2,\ldots\}$ in which the control input is updated. Thus, the closed-loop system with the neural-network feedback controller is given by:
\begin{align}\label{eq:clsys}
      \dot{x}(t)=f(x(t),N(x(t_j)),w(t)) :=  f^c(j,x(t),w(t)), 
\end{align}
where $t\in \real_{\ge 0}$ and $j\in \mathbb{Z}_{\ge 0}$ is such that $t\in [t_j,t_{j+1}]$. In our analysis when $j$ is clear from context or does not affect the result, we drop $j$ as an argument of $f^c$.

The goal of this paper is to verify the behavior of the closed-loop system \eqref{eq:clsys}. 
To verify the safety of a system under uncertainty, one needs to verify the entire reachable set. However, in general, computing the reachable set exactly is not computationally tractable---instead, approaches typically compute an over-approximation $\overline{\set{R}}_{f^{c}}(t,t_0,\set{X},\set{W})\supseteq\set{R}_{f^{c}}(t,t_0,\set{X},\set{W})$. Therefore, the main challenge addressed in this paper is to develop an approach for providing tight over-approximations of reachable sets while remaining computationally tractable for runtime computation.

\section{Interval Reachability of Neural Network Controlled Systems}

\subsection{General Framework}

We assume we have access to an off-the-shelf dynamical system reachability tool that supports \textit{interval analysis}.

\begin{assumption} [Open-loop interval reachability] \label{asm:olreach}
    Given a dynamical system of the form \eqref{eq:nlsys}, any intervals $\set{X}_0=[\underline{x}_0,\overline{x}_0]\subseteq\R^n$, $\set{U}=[\underline{u},\overline{u}]\subseteq\R^p$, and $\set{W}=[\underline{w},\overline{w}]\subseteq\R^q$, and some initial time $t_0$, there exists a reachability algorithm that returns a valid interval approximation $[\underline{x}(t),\overline{x}(t)]$ satisfying
    \begin{align} \label{eq:olreach}
    \set{R}_f(t,t_0,\set{X}_0,\set{U},\set{W}) \subseteq [\underline{x}(t),\overline{x}(t)], \quad \forall t\ge t_0.
    \end{align}
\end{assumption}

Reachability analysis of dynamical systems is a classical and well-studied research field with several off-the-shelf toolboxes with this capability, including Flow*~\cite{XC-EA-SS:13}, Hamilton-Jacobi approach~\cite{SB-MC-SH-CJT:17},, CORA~\cite{MA:15}, and mixed monotonicity~\cite{SC:20}.

We also assume access to off-the-shelf neural network verification algorithms that can provide \textit{interval inclusion functions} as follows:

\begin{assumption} [Neural network verification] \label{asm:nnverif}
    Given a neural network $N$ of the form \eqref{eq:NN} and any interval $[\underline{y},\overline{y}] \subseteq \R^n$, there exists a neural network verification algorithm that returns a valid inclusion function $\left[\begin{smallmatrix}\underline{N}_{[\underline{y},\overline{y}]}\\ \overline{N}_{[\underline{y},\overline{y}]}\end{smallmatrix}\right]: \set{T}^{2n}_{\geq 0}\to \set{T}^{2p}_{\geq 0}$ satisfying
    \begin{align}\label{eq:NNverifier}
        \underline{N}_{[\underline{y},\overline{y}]}(\underline{x},\overline{x}) \leq N(x) \leq \overline{N}_{[\underline{y},\overline{y}]}(\underline{x},\overline{x}),
    \end{align}
    for any $x\in[\underline{x},\overline{x}]\subseteq[\underline{y},\overline{y}]$.
\end{assumption}

A large number of the existing neural network verification algorithms can provide bounds of the form~\eqref{eq:NNverifier} for the output of the neural networks, including CROWN (and its subsequent variants)~\cite{HZ-etal:18}, LipSDP \cite{MF-MM-GJP:22}, and IBP \cite{SG-etal:19}.

Combining Assumptions \ref{asm:olreach} and \ref{asm:nnverif} leads naturally to an algorithm for over-approximating solutions to \eqref{eq:clsys}. In particular, starting with $j=0$ and an initial interval of states $\mathcal{X}_j=[\underline{x}(t_j),\overline{x}(t_j)]$, first obtain $\underline{N}_{[\underline{y},\overline{y}]}$ and $\overline{N}_{[\underline{y},\overline{y}]}$
for some $[\underline{y},\overline{y}]\supseteq \mathcal{X}_j$ from Assumption \ref{asm:nnverif}. Next, set $\mathcal{U}=[\underline{N}_{[\underline{y},\overline{y}]}(\underline{x}(t_j),\overline{x}(t_j)), \overline{N}_{[\underline{y},\overline{y}]}(\underline{x}(t_j),\overline{x}(t_j))]$. Finally, compute the reachable set of the closed-loop system on the interval $t\in(t_j,t_{j+1}]$ as the interval reachable set obtained from the algorithm of Assumption \ref{asm:olreach}, set $\mathcal{X}_{j+1}$ as the reachable set at time $t_{j+1}$, increment $j\gets j+1$, and iterate. This iteration serves as the backbone of our proposed paritioning-based algorithm below.

\subsection{Mixed Monotone Framework}

One specific framework that satisfies both Assumptions \ref{asm:olreach} and \ref{asm:nnverif} is developed in \cite{SJ-AH-SC:23}, where interval reachability of neural network controlled systems is studied through the lens of mixed monotone dynamical systems theory. Suppose that we have access to a decomposition function $d$ for the open-loop system~\eqref{eq:nlsys} with the open-loop embedding system
\begin{gather}\label{eq:opembsys}
    \frac{d}{dt}\begin{bmatrix}\underline{x} \\ \overline{x}\end{bmatrix} = 
    \begin{bmatrix} d(\underline{x},\overline{x},\underline{u},\overline{u},\underline{w},\overline{w}) \\ d(\overline{x},\underline{x},\overline{u},\underline{u},\overline{w},\underline{w}) \\ \end{bmatrix} := E(\underline{x},\overline{x},\underline{u},\overline{u},\underline{w},\overline{w}).
\end{gather}
Given $\set{X}_0=[\underline{x}_0,\overline{x}_0]\subseteq\R^n$, $\set{U}=[\underline{u},\overline{u}]\subseteq\R^p$, and $\set{W}=[\underline{w},\overline{w}]\subseteq\R^q$, the trajectory of \eqref{eq:opembsys} starting from $(\underline{x}_0,\overline{x}_0)$ provides the inclusion \eqref{eq:olreach} from Assumption \ref{asm:olreach}.

Following the treatment in~\cite{SJ-AH-SC:23}, one can use CROWN~\cite{HZ-etal:18} to obtain the desired bounds from Assumption \ref{asm:nnverif}. Given an interval $[\underline{y},\overline{y}]$, the algorithm provides an efficient procedure for finding a tuple $(\underline{C},\overline{C},\underline{d},\overline{d})$ defining linear upper and lower bounds for the output of the neural network
\begin{gather} \label{eq:crown}
    \underline{C}(\underline{y},\overline{y})x + \underline{d}(\underline{y},\overline{y}) \leq N(x) \leq \overline{C}(\underline{y},\overline{y})x + \overline{d}(\underline{y},\overline{y}),
\end{gather}
for every $x\in [\underline{y},\overline{y}]$. Using these linear bounds, we can construct the inclusion function for any $[\underline{x},\overline{x}]\subseteq[\underline{y},\overline{y}]$:
\begin{gather} \label{eq:crownif}
    \begin{gathered}
        \underline{N}_{[\underline{y},\overline{y}]}(\underline{x},\overline{x}) = [\underline{C}(\underline{y},\overline{y})]^+\underline{x} + [\underline{C}(\underline{y},\overline{y})]^-\overline{x} + \underline{d}(\underline{y},\overline{y}),\\
        \overline{N}_{[\underline{y},\overline{y}]}(\underline{x},\overline{x}) = [\overline{C}(\underline{y},\overline{y})]^+\overline{x} + [\overline{C}(\underline{y},\overline{y})]^-\underline{x} + \overline{d}(\underline{y},\overline{y}).
    \end{gathered}
\end{gather}

By combining these two tools, one can construct a new closed-loop embedding system to over-approximate the reachable sets of the closed-loop system \eqref{eq:clsys}. From \cite{SJ-AH-SC:23}, with the following definitions, 
\begin{align}
\begin{aligned}
   \underline{\eta}_j &= \underline{N}_{[\underline{y},\overline{y}]}(\underline{x}(t_j),\overline{x}(t_j)_{[i:\underline{x}(t_j)]}),\\ 
   \overline{\eta}_j &= \overline{N}_{[\underline{y},\overline{y}]}(\underline{x}(t_j),\overline{x}(t_j)_{[i:\underline{x}(t_j)]}),\\
   \underline{\nu}_j &= \underline{N}_{[\underline{y},\overline{y}]}(\overline{x}(t_j),\underline{x}(t_j)_{[i:\overline{x}(t_j)]}), \\
   \overline{\nu}_j &= \overline{N}_{[\underline{y},\overline{y}]}(\overline{x}(t_j),\underline{x}(t_j)_{[i:\overline{x}(t_j)]}),
\end{aligned}
\end{align}
we use the following ``hybrid'' function
\begin{align}\label{eq:closedloop-decom}
\begin{aligned}
    &\big(d_{[\underline{y},\overline{y}]}^{\mathrm{c}}(j,\underline{x},\overline{x},\underline{w},\overline{w})\big)_i  = \begin{cases}
        \begin{aligned}
            d_i (\underline{x},\overline{x}, \underline{\eta}_j, \overline{\eta}_j,\underline{w},\overline{w}), 
        \end{aligned} & \underline{x} \leq \overline{x} \\
        \begin{aligned}
            d_i (\overline{x},\underline{x}, \overline{\nu}_j, \underline{\nu}_j,\overline{w},\underline{w}), 
        \end{aligned} & \overline{x} \leq \underline{x}
    \end{cases} 
\end{aligned}
\end{align}
for any $[\underline{y},\overline{y}] \supseteq [\underline{x},\overline{x}]$, to create the closed-loop embedding system:
\begin{gather}\label{eq:clembsys}
    \frac{d}{dt}\begin{bmatrix}\underline{x} \\ \overline{x}\end{bmatrix} = 
    \begin{bmatrix} d_{[\underline{y},\overline{y}]}^{\mathrm{c}}(j,\underline{x},\overline{x},\underline{w},\overline{w}) \\ d_{[\underline{y},\overline{y}]}^{\mathrm{c}}(j,\overline{x},\underline{x},\overline{w},\underline{w}) \\ \end{bmatrix} := E_{[\underline{y},\overline{y}]}^{\mathrm{c}}(j,\underline{x},\overline{x},\underline{w},\overline{w}).
\end{gather}
In our analysis when $j$ is clear from context or does not affect the result, we drop $j$ as an argument of $d^c$ and $E^c$. For an input set $\set{X}_0\subseteq[\underline{x}_0,\overline{x}_0]$ and a disturbance set $\set{W}\subseteq[\underline{w},\overline{w}]$, with $\big[\begin{smallmatrix} \underline{x}(t) \\ \overline{x}(t) \end{smallmatrix}\big] = \phi_{E_{[\underline{y},\overline{y}]}^c}\left(t,t_0,\big[\begin{smallmatrix} \underline{x}_0 \\ \overline{x}_0 \end{smallmatrix}\big],\big[\begin{smallmatrix} \underline{w} \\ \overline{w} \end{smallmatrix}\big]\right)$ for every $t>t_0$, the following inclusion holds~\cite[Theorem 1]{SJ-AH-SC:23}:
\begin{gather}\label{eq:overapp}
    \set{R}_{f^c}(t,t_0,\set{X},\set{W}) \subseteq [\underline{x}(t),\overline{x}(t)].
\end{gather}
Thus, running one single trajectory of $E^{c}$ can over-approximate the reachable set of the closed-loop system~\eqref{eq:clsys}. 

\subsection{Partitioning}

While interval analysis techniques can often be computationally inexpensive, as the size of the uncertain sets grow, the reachable set estimates become overly conservative. In a closed-loop, these effects can compound significantly as the error continues to accumulate. This phenomenon is known in the literature as the \textit{wrapping effect}. To mitigate over-conservatism in interval-based techniques, one can split uncertain regions into smaller partitions. A valid over-approximation of the reachable set can be found by taking the union of the over-approximations of the reachable sets for each partition. Due to the locality of the smaller partitions, this technique can drastically reduce the conservatism of interval-based techniques.

\section{Contraction-Guided Adaptive Partitioning} \label{sec:alg}

\SetKwFunction{crs}{compute\_reachable\_set}\SetKwFunction{step}{step}

In this section, we introduce a contraction-guided adaptive partitioning algorithm to improve the results of interval reachability methods satisfying Assumptions~\ref{asm:olreach} and \ref{asm:nnverif}. 
This algorithm is motivated by the theory in the next section.

\begin{algorithm} \setstretch{1.05}
\caption{Contraction-Guided Adaptive Partitioning (ReachMM-CG)} \label{alg:main}
\KwIn{initial set: $\set{X}_0=[\underline{X}_0,\overline{X}_0]\subset\R^n$, control time instances: $\{t_0,t_1,\ldots,\}$, final time: $T$}
\Parameter{desired width $\varepsilon\in\overline{\R}^n_{\geq 0}$, check contraction factor $\gamma\in(0,1]$, max partition depth $D_p\in\mathbb{Z}_{\geq 0}$, max neural network verification depth $D_{\textsc{N}}\in\mathbb{Z}_{\geq 0}$}
\KwOut{Over-approximated reachable set trajectory for $t\in [t_0,T]$: $\overline{\set{R}}(t)$}
\SetKwProg{myalg}{Procedure}{}{} \SetKwProg{myproc}{Procedure}{}{}
    $\overline{\set{R}}(t_0) = \set{X}_0$\;
    $\set{P} = ((\underline{X}_0,\overline{X}_0), \textsc{True}, \emptyset)$\;
    $t_m = $ smallest control time instance $\geq T$\; 
    \For{$j=1,\ldots,m$}{
        $(\overline{\set{R}}(\cdot)|_{[t_{j-1},t_j]}, \set{P}_+) \gets$ \step{$\set{P},\emptyset,j$}\;
        $\set{P} \gets \set{P}_+$\;
    }
    \Return $\overline{\set{R}}(t) \ \forall t\in[t_0,T]$\;
\myproc{\step{$((\underline{x}_0,\overline{x}_0),\textsc{N},\set{S}), E^c, j$}}{
    \uIf(\tcp*[f]{NN Verification}){$\textsc{N}$ is \textsc{True}} {
        $E^c \gets$ $E^c$ as (\ref{eq:clembsys}, \ref{eq:closedloop-decom}) using \eqref{eq:crownif} on $(\underline{x}_0,\overline{x}_0)$\; \label{alg:line:crown}
    }
    \uIf(\tcp*[f]{No Subpartitions}){$\set{S} = \emptyset$}{
        $d = \operatorname{get\_depth()}$ \tcp*{Tree Depth}
        \uIf(\tcp*[f]{Can Partition}){$d < D_p$}{
            $t_\gamma \gets t_{j-1} + \gamma(t_{j} - t_{j-1})$\;
            $(\underline{x},\overline{x})(t_\gamma)\gets \phi_{E^c}(t_\gamma,t_{j-1},(\underline{x}_0,\overline{x}_0))$\;
            $C = \frac{\|\overline{x}(t_\gamma)-\underline{x}(t_\gamma)\|_{\infty,\varepsilon}}{\|\overline{x}_0 - \underline{x}_0\|_{\infty,\varepsilon}}$\tcp*{Contr Rate} \label{alg:main:contrrate}
            \uIf{$C^{1/\gamma}\|\overline{x}_0 - \underline{x}_0\|_{\infty,\varepsilon} > 1$}{
                $\{\underline{x}_0^k,\overline{x}_0^k\}_{k=1}^{2^n} \gets \operatorname{uni\_div}(\underline{x}_0,\overline{x}_0)$\; \label{alg:main:unidiv}
                $\set{S} \gets \{((\underline{x}^k_0,\overline{x}^k_0), d < D_{\textsc{N}}, \emptyset)\}_{k=1}^{2^n}$\;
                $\textsc{N} \gets \textsc{N} \land ((d+1) > D_{\textsc{N}})$\;
                \Return \step{$((\underline{x}_0,\overline{x}_0),\textsc{N},\set{S}),E^c,j$}
            } 
        }
        $(\underline{x},\overline{x})(\cdot)|_{[t_{j-1},t_j]} \gets \phi_{E^c}(\cdot,t_{j-1},(\underline{x}_0,\overline{x}_0))$\;
        \Return $([\underline{x},\overline{x}](\cdot)|_{[t_{j-1},t_j]}, ((\underline{x},\overline{x})(t_f),\textsc{N},\emptyset))$
    } \uElse (\tcp*[f]{Iterate Subpartitions}) {
        $\{(\overline{\set{R}}^k(\cdot), P_+^k) \gets $\step{$P^k,E^c,j$}$\}_{P^k\in \set{S}}$\;
        $(\underline{x}_0,\overline{x}_0)_+ \gets \text{tightest } [\underline{x},\overline{x}] \supseteq \bigcup_k \overline{\set{R}}^k(t_f)$\;
        \Return $(\bigcup_k \overline{\set{R}}^k(\cdot), ((\underline{x}_0,\overline{x}_0)_+, \textsc{N}, \{P_+^k\}_{k=1}^{2^n}))$\;
    }
}
\end{algorithm}

\subsection{Algorithm Description}
At a particular control step $t_j$, define a partition
\begin{align}
    P = ((\underline{x}_0,\overline{x}_0),\text{N},\set{S}), \label{eq:partition}
\end{align}
where $(\underline{x}_0,\overline{x}_0)\in\set{T}^{2n}_{\geq 0}$ is the initial condition of the partition, $\text{N}\in \{\textsc{True},\textsc{False}\}$ is whether or not the partition should compute the neural network verification (\ref{eq:NNverifier}), and $\set{S}$ is a set of tuples representing its subpartitions.  Algorithm~\ref{alg:main} starts by initializing a tree at time $t_0$ with a single root node $\set{P}=((\underline{X}_0,\overline{X}_0),\textsc{True},\emptyset)$ representing the initial set. 

\paragraph*{\step{} Procedure}
Along a particular control interval $[t_{j-1},t_j]$, every partition undergoes the following procedure. If specified, the embedding system $E^c$ is computed using (\ref{eq:crownif}, \ref{eq:closedloop-decom}, \ref{eq:clembsys}). Initially, $E^c$ is integrated to a fraction $\gamma\in(0,1]$ of the control interval, where the contraction rate is computed by comparing its width to its initial condition. Then, the width of the final box at $t_j$ is estimated. If it violates hyper-parameter $\varepsilon$, then the algorithm returns to time $t_{j-1}$ and divides the initial condition into $2^n$ sub-partitions. This process is repeated for each of the sub-partitions. Finally, if the estimated width does not violate $\varepsilon$, $E^c$ is fully integrated to $t_j$. This procedure is visualized in Figure~\ref{fig:step}, and formalized as \step{} in Algorithm~\ref{alg:main}, which repeats for every control step until some final time $T$.

\paragraph*{Hyper-parameters}
There are several user-defined parameters that are important for the algorithm's performance. The choice of the check-contraction factor $\gamma\in(0,1]$ is purely for computational benefit, and in particular, $\gamma=1$ checks the true width of the box at $t_j$ instead of an estimate. The maximum partition depth $D_p\in\mathbb{Z}_{\geq 0}$ specifies how deep new partitions can be added in the tree to integrate $E^c$. The maximum neural network verification depth $D_{\textsc{N}}\in\mathbb{Z}_{\geq 0}$ specifies how deep neural network verification is performed in the partition tree. $D_p$ and $D_\textsc{N}$ are directly related to the first two terms of the right-hand side of Theorem~\ref{thm:main} in the next section---every partitioning layer will improve the first term, while only partitions computing the neural network verification will improve the second term. Finally, the choice of desired width $\varepsilon\in\overline{\R}^n_{\geq 0}$ specifies how wide partitions can grow before being cut: if chosen too small, the algorithm will partition early in the start ($\varepsilon_i=0$ for any $i$ implies a uniform initial partitioning); if chosen too large, the algorithm will not partition at all ($\varepsilon_i=\infty$ implies no condition on the $i$-th component).

\begin{figure}
    \label{fig:parttree}
    \centering
    \includegraphics[width=0.95\linewidth]{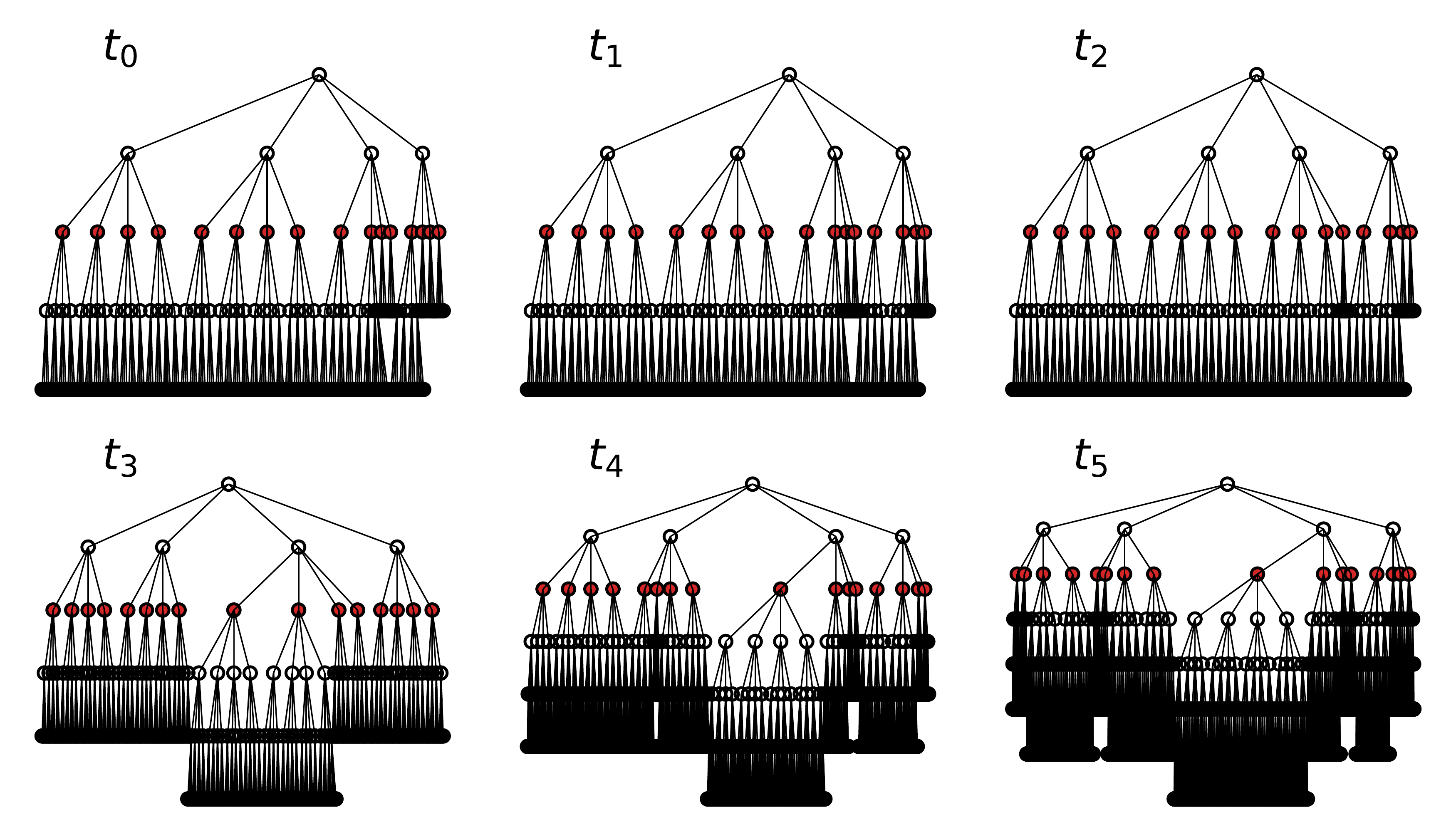}
    \caption{A partition tree structure of Algorithm~\ref{alg:main} for a run on the double integrator system~\eqref{eq:doubleintegrator} for $\varepsilon=0.1$, $D_p=10$, $D_\text{N}=2$. The color represents the algorithm's \textit{separation}---only nodes filled with {\color{tab:red} red} compute the neural network verification step, and all the integrations are performed only on the leaf nodes. The imbalanced structure is a consequence of the algorithm's \textit{spatial awareness}. As a consequence of the algorithm's \textit{temporal awareness}, the structure of the partition tree deepens as time increases.}
    \vspace{-1em}
\end{figure}

\subsection{Discussion}
Algorithm~\ref{alg:main} has three key features.

\paragraph*{Separation}
In practice, the neural network verification step usually contributes the most computational expense to the algorithm. Introducing the maximum depths $D_p$ and $D_\text{N}$ allows us to separate calls to the neural network verification algorithm and the total number of partitions. This allows the algorithm to improve its accuracy without any additional calls to the neural network verifier, leaving it to the user to control the trade-off between computational complexity and accuracy. In Theorem~\ref{thm:main}, this corresponds with an improvement of the first term separately from the second term.

\paragraph*{Spatial awareness}
Based on the dynamics of the system, the over-approximation error can be highly sensitive to the spatial location of each partition. By estimating the contraction rate of each partition separately, the algorithm can localize partitioning exactly \textit{where} necessary. In Theorem~\ref{thm:main}, this curbs exponential growth \textit{where} $c_x$ is the worst.

\paragraph*{Temporal awareness}
One of the main drawbacks of interval reachability frameworks is that as time evolves, the over-approximation error tends to compound exponentially. As such, by partitioning along trajectories, the algorithm can save computations to improve reachable set estimates exactly \textit{when} necessary. In Theorem~\ref{thm:main}, this curbs exponential growth \textit{when} $c_x$ is the worst.

\paragraph*{Generality of Algorithm \ref{alg:main}}
The algorithm, as written, specifically uses the Mixed Monotone framework (developed in Section V.B) for the interval reachability of the dynamical system, as well as CROWN (specified by \eqref{eq:crown}) for the neural network verification. We will exclusively use this setting for the rest of the paper, and refer it as ``ReachMM-CG''. However, it is easy to see that one can replace Lines 11, 16, and 23 with any other interval reachability setting satisfying Assumptions \ref{asm:olreach} and \ref{asm:nnverif}.

\section{Contraction-Based Guarantees on Reachable Set Over-Approximation} \label{sec:theory}

In this section, we use contraction theory to provide rigorous bounds on the accuracy of the mixed-monotone reachable set over-approximations. For the simplicity of analysis in this section, we remove the assumption that the control is applied in the piecewise constant fashion and instead consider a continuously applied neural network.
\vspace{-0.2cm}
\begin{theorem}[Accuracy guarantees for mixed-monotone reachability]\label{thm:main}
Consider the closed-loop system~\eqref{eq:clsys} with the neural network controller $u=N(x)$ given by~\eqref{eq:NN} and let $t\mapsto x(t)$ be a trajectory of the closed-loop system~\eqref{eq:clsys}. Suppose that $d$ is the decomposition function for the open-loop system~\eqref{eq:nlsys}, there is a neural network verification algorithm which provides the bounds of the form~\eqref{eq:NNverifier}. Let $t\mapsto \left[\begin{smallmatrix}\underline{x}(t)\\ \overline{x}(t)\end{smallmatrix}\right]$ be a trajectory of the closed-loop embedding system~\eqref{eq:clembsys} and $t\mapsto \left[\begin{smallmatrix}\underline{y}(t)\\ \overline{y}(t)\end{smallmatrix}\right]$ be a curve such that $[\underline{x}(t),\overline{x}(t)] \subseteq [\underline{y}(t),\overline{y}(t)]$, for every $t\in \real_{\ge 0}$. Then, with $t_0=0$,
    \begin{multline}\label{eq:bound}
    \left\|\begin{bmatrix}\underline{x}(t)\\ \overline{x}(t)\end{bmatrix} - \begin{bmatrix}x(t)\\ x(t)\end{bmatrix}\right\|_{\infty} \le e^{c_x t} \left\|\begin{bmatrix}\underline{x}(0)\\ \overline{x}(0)\end{bmatrix} - \begin{bmatrix}x(0)\\ x(0)\end{bmatrix}\right\|_{\infty} \\ + \frac{\ell^o_u (e^{c_x t}-1)}{c_x} \sup_{\substack{\tau\in [0,t],\\ z\in [\underline{y}(\tau),\overline{y}(\tau)]}}\left\|\begin{bmatrix}\underline{N}_{[\underline{y},\overline{y}]}(z,z)\\ \overline{N}_{[\underline{y},\overline{y}]}(z,z)\end{bmatrix} - \begin{bmatrix}N(z)\\ N(z)\end{bmatrix}\right\|_{\infty} \\ + \frac{\ell^{\mathrm{o}}_w (e^{c_x t}-1)}{c_x} \sup_{\tau\in [0,t]}\left\|\begin{bmatrix}\underline{w}(\tau)\\ \overline{w}(\tau)\end{bmatrix} - \begin{bmatrix}w(\tau)\\ w(\tau)\end{bmatrix}\right\|_{\infty}
    \end{multline}
    where
    \begin{align*}
       c_x &= \sup_{\underline{z},\overline{z}\in \Omega_t}\mu_{\infty}(D_{\left[\begin{smallmatrix}\underline{z}\\\overline{z}\end{smallmatrix}\right]}E^{\mathrm{c}}_{[\underline{y}(t),\overline{y}(t)]}(\underline{z},\overline{z},\underline{w},\overline{w})),\\
       \ell^{\mathrm{o}}_u &= \sup_{\substack{x\in \Omega_t, w \in [\underline{w},\overline{w}],\\ \underline{u},\overline{u}\in [\underline{N}(x,x),\overline{N}(x,x)]}}\|D_{\left[\begin{smallmatrix}\underline{u}\\\overline{u}\end{smallmatrix}\right]} E^{\mathrm{o}}(x,x,\underline{u},\overline{u},w,w)\|_{\infty},\\
       \ell^{\mathrm{o}}_w &= \sup_{\substack{x\in \Omega_t, \underline{z},\overline{z}\in [\underline{w},\overline{w}],\\ \underline{u},\overline{u}\in [\underline{N}(x,x),\overline{N}(x,x)]}}\|D_{\left[\begin{smallmatrix}\underline{z}\\\overline{z}\end{smallmatrix}\right]} E^{\mathrm{o}}(x,x,\underline{u},\overline{u},\underline{z},\overline{z})\|_{\infty},   
    \end{align*}
    and the set $\Omega_t$ is defined by $\Omega_t = \bigcup_{\tau\in [0,t]}[\underline{y}(\tau),\overline{y}(\tau)]\subseteq \real^n$. 
\end{theorem}
\vspace{-2em}
\begin{proof}
We first make the following conventions:
\begin{align*}
r(t) & = \begin{bmatrix}\underline{x}(t)\\ \overline{x}(t)\end{bmatrix} - \begin{bmatrix}x(t)\\ x(t)\end{bmatrix}, \\
    \textcircle{1}&= E^c_{[\underline{y},\overline{y}]}(\underline{x},\overline{x},\underline{w},\overline{w})-E_{[\underline{y},\overline{y}]}^c(x,x,\underline{w},\overline{w}),\\
    \textcircle{2} &= E^c_{[\underline{y},\overline{y}]}(x,x,\underline{w},\overline{w})-E_{[\underline{y},\overline{y}]}^c(x,x,w,w), \\
    \textcircle{3} &= E^c_{[\underline{y},\overline{y}]}(x,x,w,w)-\begin{bmatrix} f^c(x,w)\\ f^c(x,w)\end{bmatrix}.
\end{align*}
Therefore, we can compute 
\begin{multline*}
 \tfrac{1}{2}D^{+} \|r(t)\|^2_{\infty} = \WP{E^c_{[\underline{y},\overline{y}]}(\underline{x},\overline{x},\underline{w},\overline{w})-\left[\begin{smallmatrix}
    f^c(x,w)\\ f^c(x,w)
    \end{smallmatrix}\right]}{r(t)}_{\infty} \\ \le \WP{\textcircle{1}}{r(t)}_{\infty}  + 
 \WP{\textcircle{2}}{r(t)}_{\infty}+\WP{\textcircle{3}}{r(t)}_{\infty},
\end{multline*}
where $D^{+}$ is the upper Dini Derivative with respect to time, the first equality holds by the curve-norm derivative property of the weak pairing $\WP{\cdot}{\cdot}_{\infty}$~\cite[Theorem 25(ii)]{AD-SJ-FB:20o} and the second inequality holds by the subadditivity of weak pairing $\WP{\cdot}{\cdot}_{\infty}$~\cite[Definition 15, property (i)]{AD-SJ-FB:20o}. By~\cite[Theorem 18]{AD-SJ-FB:20o}, the first term on the RHS of the above equation can be estimated $\WP{\textcircle{1}}{r(t)}_{\infty} \le c_x \|r(t)\|^2_{\infty}$. We also note that, for the closed-loop system, we have
\begin{align*}
  &E^{\mathrm{c}}_{[\underline{y},\overline{y}]}(x,x,\underline{w},\overline{w}) = E^\mathrm{o}(x,x,\underline{N}_{[\underline{y},\overline{y}]}(x,x),\overline{N}_{[\underline{y},\overline{y}]}(x,x),\underline{w},\overline{w}),\\
  &E^{\mathrm{c}}_{[\underline{y},\overline{y}]}(x,x,w,w) = E^\mathrm{o}(x,x,\underline{N}_{[\underline{y},\overline{y}]}(x,x),\overline{N}_{[\underline{y},\overline{y}]}(x,x),w,w).
\end{align*}
Therefore, by~\cite[Definition 15, property (iv)]{AD-SJ-FB:20o}, the second term in the RHS can be estimated as 
\begin{align*}
   \WP{\textcircle{2}}{r(t)}_{\infty} \le \ell^{\mathrm{o}}_w\left\|\begin{bmatrix}\underline{w}(t)\\ \overline{w}(t)\end{bmatrix} - \begin{bmatrix}w(t)\\ w(t)\end{bmatrix}\right\|_{\infty}\|r(t)\|_{\infty}.
\end{align*}
Finally, by the definition of $E^c$, we have
\begin{align*}
\textcircle{3} & = E^{o}(x,x,\underline{N}_{[\underline{y},\overline{y}]}(x,x),\overline{N}_{[\underline{y},\overline{y}]}(x,x),w,w) \\ & - E^o(x,x,N(x),N(x),w,w). 
\end{align*}
Thus, by~\cite[Definition 15, property (iv)]{AD-SJ-FB:20o}, the third term in the RHS can be estimated as
\begin{align*}
   \WP{\textcircle{3}}{r(t)}_{\infty} \le \ell^o_u\left\|\begin{bmatrix}\underline{N}_{[\underline{y},\overline{y}]}(x,x)\\ \overline{N}_{[\underline{y},\overline{y}]}(x,x)\end{bmatrix} - \begin{bmatrix}N(x)\\ N(x)\end{bmatrix}\right\|_{\infty}\|r(t)\|_{\infty}. 
\end{align*}
Therefore, gathering all the terms and dividing by $\|r(t)\|_\infty$,
\begin{align*}
    D^{+}\|r(t)\|_{\infty} & \le c_x \|r(t)\|_{\infty}   + \ell^o_u \left\|\begin{bmatrix}\underline{N}_{[\underline{y},\overline{y}]}(x,x)\\ \overline{N}_{[\underline{y},\overline{y}]}(x,x)\end{bmatrix} - \begin{bmatrix}N(x)\\ N(x)\end{bmatrix}\right\|_{\infty}  \\ & + \ell^{\mathrm{o}}_{w} \left\|\begin{bmatrix}\underline{w}(t)\\ \overline{w}(t)\end{bmatrix} - \begin{bmatrix}w(t)\\ w(t)\end{bmatrix}\right\|_{\infty}. 
\end{align*}
The result then follows from the Gr\"{o}nwall\textendash{}Bellman inequality~\cite[Lemma 11]{AD-SJ-FB:20o}.
\end{proof}
\begin{remark}
\;
\begin{enumerate}
    \item Theorem~\ref{thm:main} explains the three key features of Algorithm~\ref{alg:main}. (i) separation allows the improvement of term 1 independent of term 2 on the RHS of \eqref{eq:bound}; (ii) spatial awareness allows for cutting the partitions with the worst contraction rates $c_x$ and lipschitz bounds $\ell_u^o,\ell_w^o$, since each partition defines a different set $\Omega_t$; (iii) temporal awareness allows for improving the exponential growth present in $e^{c_xt}$. The hyper-parameters $D_p,D_\text{N}$ of Algorithm~\ref{alg:main} control the first and second terms respectively of \eqref{eq:bound}.

\item The constant $c_x$ in Theorem~\ref{thm:main} is the contraction rate of the closed-loop embedding system~\eqref{eq:clembsys}. Theorem~\ref{thm:main} shows that $c_x$ is also the rate of contraction of the box over-approximation of reachable sets using mixed monotonicity. For $c_x<0$, the upper bound shows that the error associated with uncertainties in initial conditions converges to $0$ as $t\to \infty$.

\end{enumerate}
\end{remark}

Finding the value of the contraction parameter $c_x$ in Theorem~\ref{thm:main} is usually complicated due to the large number of parameters of the neural network and its interconnection with the nonlinear dynamics of the system. In the next theorem, we provide a simple method for over-approximating $c_x$ using the input-output Lipschitz bounds of the neural network.

\begin{theorem}[Upper bound on closed-loop contraction rate] \label{thm:clcont}
    Consider the closed-loop embedding system~\eqref{eq:clembsys} with the neural network controller $u=N(x)$ given by~\eqref{eq:NN}. Suppose that, for the neural network verification in~\eqref{eq:NNverifier}, we have
    \begin{align*}  \sup_{x\in \Omega_t}\left\|D_{\left[\begin{smallmatrix}\underline{x}\\\overline{x}\end{smallmatrix}\right]}\left[\begin{smallmatrix}\underline{N}_{[\underline{y},\overline{y}]}\\\overline{N}_{[\underline{y},\overline{y}]}\end{smallmatrix}\right]\right\|_{\infty} = \mathrm{Lip}_{\infty}
    \end{align*}
    and, for the open-loop embedding system~\eqref{eq:opembsys}, we have 
    \begin{align*}
     c_x^{\mathrm{o}} &= \sup_{\substack{\underline{x},\overline{x}\in \Omega_t, w \in [\underline{w},\overline{w}],\\ \underline{u},\overline{u}\in [\underline{N}(\underline{x},\overline{x}),\overline{N}(\underline{x},\overline{x})]}}\left(\mu_{\infty}(D_{\left[\begin{smallmatrix}\underline{x}\\\overline{x}\end{smallmatrix}\right]}E^{\mathrm{o}}(\underline{x},\overline{x},\underline{u},\overline{u},\underline{w},\overline{w}))\right)\\
     \ell_u^{\mathrm{o}} &= \sup_{\substack{\underline{x},\overline{x}\in \Omega_t, w \in [\underline{w},\overline{w}],\\ \underline{u},\overline{u}\in [\underline{N}(\underline{x},\overline{x}),\overline{N}(\underline{x},\overline{x})]}}\left\|D_{\left[\begin{smallmatrix}\underline{u}\\\overline{u}\end{smallmatrix}\right]}E^{\mathrm{o}}(\underline{x},\overline{x},\underline{u},\overline{u},\underline{w},\overline{w})\right\|_{\infty}
    \end{align*}
    where the set $\Omega_t$ is defined by $\Omega_t = \bigcup_{\tau\in [0,t]}[\underline{y}(\tau),\overline{y}(\tau)]\subseteq \real^n$. Then $\mu_{\infty}(D_{\left[\begin{smallmatrix}\underline{x}\\\overline{x}\end{smallmatrix}\right]}E_{[\underline{y},\overline{y}]}^{\mathrm{c}}) \le c_x^{\mathrm{o}} + \ell_u^{\mathrm{o}} \mathrm{Lip}_{\infty}$. 
\end{theorem}
\vspace{-1em}
\begin{proof}
Note that, for every $\underline{x}\le \overline{x}$ and every $\underline{w}\le \overline{w}$,
\begin{align*}
    E^{c}_{[\underline{y},\overline{y}]}(\underline{x},\overline{x},\underline{w},\overline{w}) = E^o(\underline{x},\overline{x},\underline{N}_{[\underline{y},\overline{y}]}(\underline{x},\overline{x}), \overline{N}_{[\underline{y},\overline{y}]}(\underline{x},\overline{x}),\underline{w},\overline{w}).
\end{align*}
    Therefore, using the chain rule, we have
    \begin{align*}
        &\mu_\infty\left(D_{\left[\begin{smallmatrix}\underline{x}\\\overline{x}\end{smallmatrix}\right]} E_{[\underline{y},\overline{y}]}^{c}\right) = \mu_\infty\left(D_{\left[\begin{smallmatrix}\underline{x}\\\overline{x}\end{smallmatrix}\right]} E^{o} + D_{\left[\begin{smallmatrix}\underline{u}\\\overline{u}\end{smallmatrix}\right]} E^{o} D_{\left[\begin{smallmatrix}\underline{x}\\\overline{x}\end{smallmatrix}\right]}\left[\begin{smallmatrix}\underline{N}_{[\underline{y},\overline{y}]}\\\overline{N}_{[\underline{y},\overline{y}]}\end{smallmatrix}\right]\right) \\
        &\leq 
        \mu_\infty\left(D_{\left[\begin{smallmatrix}\underline{x}\\\overline{x}\end{smallmatrix}\right]} E^{o}\right) +\mu_{\infty}\left( D_{\left[\begin{smallmatrix}\underline{u}\\\overline{u}\end{smallmatrix}\right]} E^{o} \;\;D_{\left[\begin{smallmatrix}\underline{x}\\\overline{x}\end{smallmatrix}\right]}\left[\begin{smallmatrix}\underline{N}_{[\underline{y},\overline{y}]}\\\overline{N}_{[\underline{y},\overline{y}]}\end{smallmatrix}\right]\right) \\ & \leq \mu_\infty\left(D_{\left[\begin{smallmatrix}\underline{x}\\\overline{x}\end{smallmatrix}\right]} E^{o}\right) +\left\| D_{\left[\begin{smallmatrix}\underline{u}\\\overline{u}\end{smallmatrix}\right]} E^{o} \right\|_{\infty} \left\|D_{\left[\begin{smallmatrix}\underline{x}\\\overline{x}\end{smallmatrix}\right]}\left[\begin{smallmatrix}\underline{N}_{[\underline{y},\overline{y}]}\\\overline{N}_{[\underline{y},\overline{y}]}\end{smallmatrix}\right]\right\|_{\infty} \\ & \le  c^{o}_x + \ell^{o}_u \mathrm{Lip}_{\infty}.       
    \end{align*}
    where the second inequality holds by the subadditive property of matrix measure $\mu_{\infty}(\cdot)$~\cite[Lemma 2.10]{FB:22-CTDS} and the third inequality holds using the fact that $\mu_{\infty}(AB)\le \|AB\|_{\infty}\le \|A\|_{\infty}\|B\|_{\infty}$, for every $A\in \real^{2n\times 2p}$ and $B\in \real^{2p\times 2n}$.
\end{proof}

\section{Numerical Experiments}

\subsection{Nonlinear vehicle} \let\thefootnote\relax\footnote{All code is available at \newline\url{https://github.com/gtfactslab/ReachMM_CDC2023}}
Consider the nonlinear dynamics of a vehicle from~\cite{PP-FA-BdAN-AdlF:17}:
\begin{align}
\begin{aligned} \label{eq:vehicle}
     \dot{p_x} &= v \cos(\phi + \beta(u_2)) &\ \dot{\phi} &=\frac{v}{\ell_r}\sin(\beta(u_2))\\
     \dot{p_y} &= v \sin(\phi + \beta(u_2)) &\ \dot{v} &= u_1
\end{aligned} 
\end{align}
where $[p_x,p_y]^{\top}\in \real^2$ is the displacement of the center of mass, $\phi \in [-\pi,\pi)$ is the heading angle in the plane, and $v\in \real_{\geq 0}$ is the speed of the center of mass. Control input $u_1$ is the applied force, input $u_2$ is the angle of the front wheels, and $\beta(u_2) = \mathrm{arctan}\left(\frac{\ell_f}{\ell_f+\ell_r}\tan(u_2)\right)$ is the slip slide angle. Let $x=[p_x,p_y,\phi,v]^\top$ and $u=[u_1,u_2]^\top$. We use the open-loop decomposition function and the neural network controller ($4\times 100\times 100\times 2$ ReLU) defined in \cite{SJ-AH-SC:23}. The neural network controller is applied at evenly spaced intervals of $0.25$ seconds apart. This neural network is trained to mimic an MPC that stabilizes the vehicle to the origin while avoiding a circular obstacle centered at $(4,4)$ with a radius of $2$.

\begin{figure}
    \centering
    \includegraphics[width=0.5\linewidth]{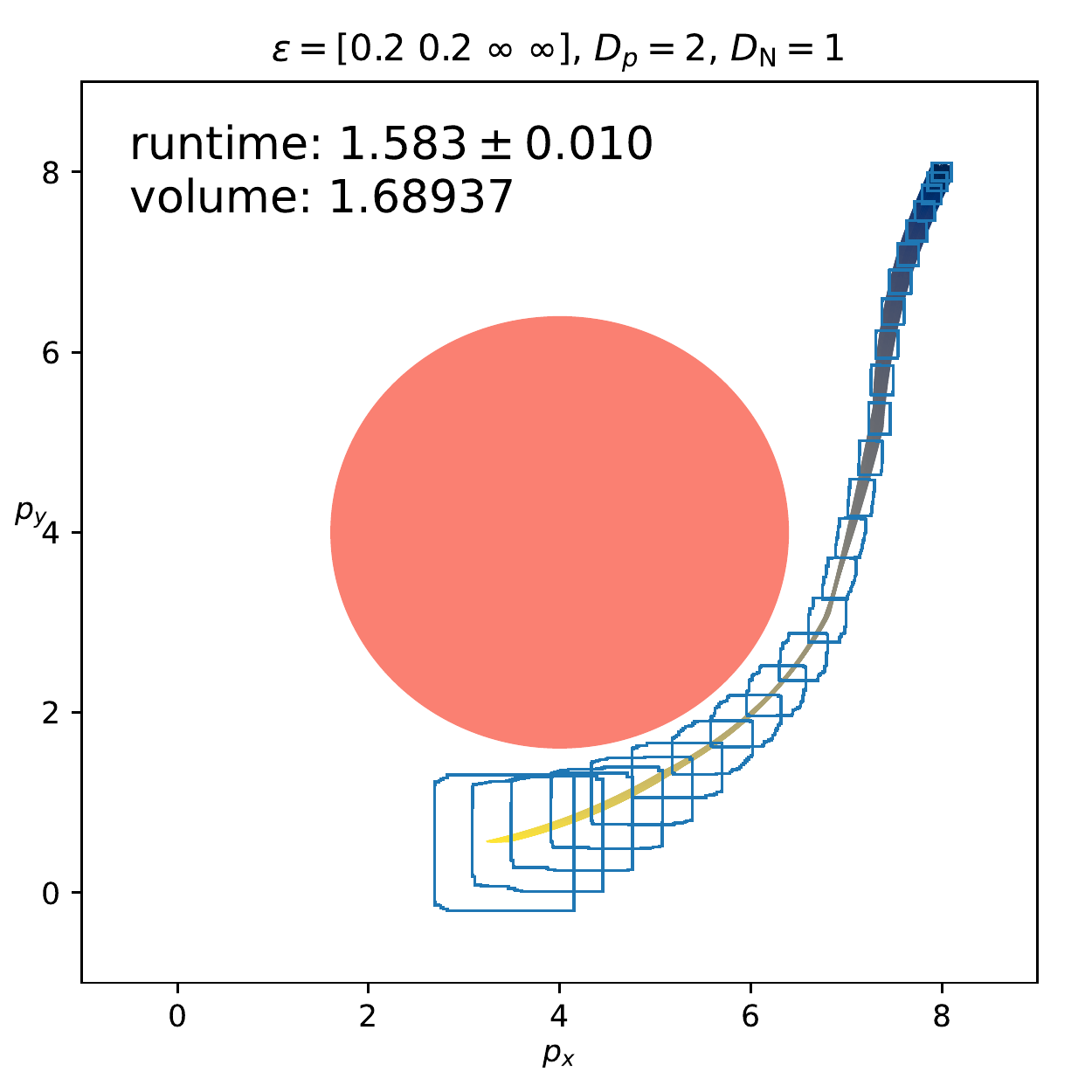}
    \caption{The over-approximated reachable sets of the closed-loop nonlinear vehicle model \eqref{eq:vehicle} in the $(p_x,p_y)$ coordinates are shown in {\color{tab:blue} blue} for the initial set $[7.9,8.1]^2\times[-\frac{2\pi}{3}-0.01, -\frac{2\pi}{3}+0.01]\times[1.99,2.01]$ over the time interval $[0,1.25]$. They are computed using Algorithm \ref{alg:main} with $\varepsilon=[0.2,0.2,\infty,\infty]^\top$, $D_p=2$, and $D_\text{N}=1$. The average runtime across 100 runs with standard deviation is reported, as well as the volume of the over-bounding box at the final time $T=1.25$. 200 true trajectories of the system are shown in the time-varying yellow line.} 
    \label{fig:vehfig1}
    \vspace{-1em}
\end{figure}

ReachMM-CG (presented in Algorithm~\ref{alg:main}) is used to provide over-approximations of the reachable sets for $t\in[0,1.25]$ as pictured in Figure \ref{fig:vehfig1}. We consider the initial set of $[7.9,8.1]^2\times[-\frac{2\pi}{3}-0.01, -\frac{2\pi}{3}+0.01]\times[1.99,2.01]$. All integrations are performed using Euler integration with a step size of $0.01$, and CROWN is computed using auto\_LiRPA~\cite{xu2020automatic}. In Table~\ref{tab:vehtable}, the performance of ReachMM-CG with different selections of hyper-parameters (all with check contraction factor $\gamma=0.1$) is compared to ReachMM, the non-adaptive purely initial set partitioning strategy from \cite{SJ-AH-SC:23}. The runtimes are averaged over 100 runs, with mean and standard deviation reported. The volumes for the tightest over-bounding box for the approximated reachable set at the final time step is reported. The adaptive strategies see an improved run-time over the non-adaptive counterparts, and in some cases see a tighter approximation.

\begin{table}[h!]
    \centering
    \begin{tabular}{||c c c c||}
        \hline
         $\varepsilon$                  & $D_p,D_\textsc{N}$ & Runtime (s)     & Volume  \\
        \hline\hline
 non-adaptive               & $(2,1)$            & $1.851\pm0.010$ & $1.988$ \\
 $[0.2,\,0.2,\,\infty,\,\infty]$ & $(2,1)$            & $1.583\pm0.010$ & $1.689$ \\
  $[0.25,\,0.25,\,\infty,\,\infty]$ & $(2,1)$            & $1.243\pm0.008$ & $1.846$ \\
 non-adaptive               & $(2,2)$            & $4.274\pm0.023$ & $0.803$ \\
 $[0.2,\,0.2,\,\infty,\,\infty]$ & $(2,2)$            & $3.332\pm0.012$ & $0.787$ \\
 $[0.25,\,0.25,\,\infty,\,\infty]$ & $(2,2)$            & $2.636\pm0.008$ & $0.986$ \\
        \hline
    \end{tabular}
    \caption{The performance of ReachMM-CG on the vehicle model. }
    \label{tab:vehtable}
\end{table}

\vspace{-1em}
\subsection{Linear Discrete-Time Double Integrator}

\begin{figure}
    \centering
    \includegraphics[width=0.95\linewidth]{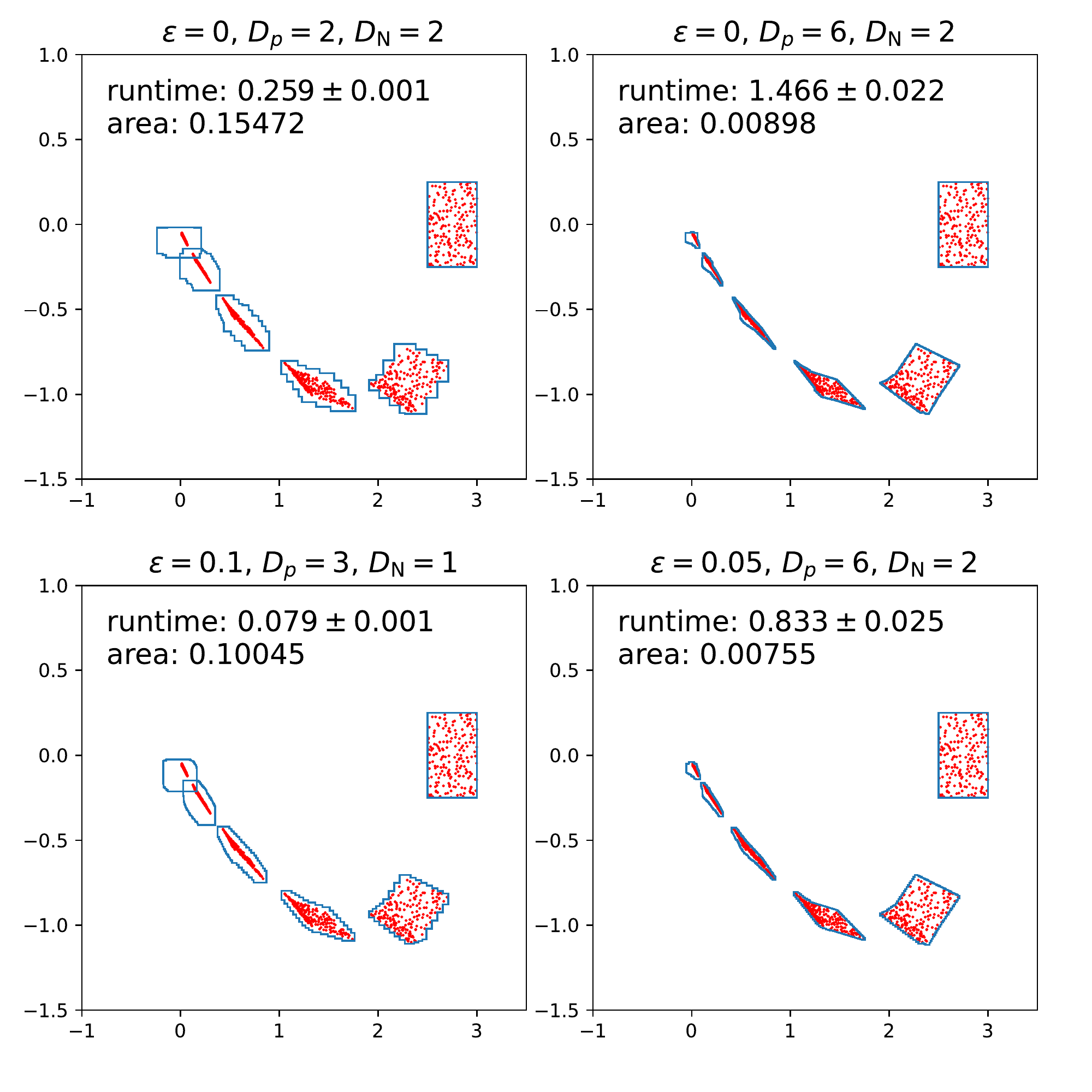}
    \caption{The over-approximated reachable sets of the closed-loop double integrator model~\eqref{eq:doubleintegrator} are shown in {\color{tab:blue} blue} for the initial set $[2.5,3]\times[-0.25,0.25]$ and final time $T=5$. They are computed using Algorithm 1 with the specified parameters in the title of each plot. The average runtime and standard deviation across 100 runs is reported, as well as the true area of the final reachable set. 200 true trajectories are shown in {\color{tab:red} red}. The horizontal axis is $x_1$ and the vertical axis is $x_2$. }
    \label{fig:DI_fig1}
    \vspace{-1em}
\end{figure}

Consider a zero-order hold discretization of the classical double integrator with a step-size of $1$ (from \cite{ME-GH-CS-JPH:21}):
\begin{align} \label{eq:doubleintegrator}
    x_{t+1} = \underbrace{\begin{bmatrix} 1 & 1 \\ 0 & 1 \end{bmatrix}}_{A} x_t + \underbrace{\begin{bmatrix} 0.5 \\ 1 \end{bmatrix}}_B u_t
\end{align}
Here, we consider a fixed actuation step size of $1$---the same as the integration step. In this special case (discrete-time LTI systems), \cite[Corollary 4]{SJ-AH-SC:23} shows that the following is a valid closed-loop embedding system that provides tighter bounds than \eqref{eq:clembsys}:
\begin{align*}
    \begin{bmatrix} \underline{x}_{t+1} \\ \overline{x}_{t+1} \end{bmatrix} = \begin{bmatrix} \underline{M}^+ & \underline{M}^- \\ \overline{M}^- & \overline{M}^+ \end{bmatrix}\begin{bmatrix} \underline{x}_{t} \\ \overline{x}_{t} \end{bmatrix} + \begin{bmatrix} B^+ & B^- \\ B^- & B^+ \end{bmatrix}\begin{bmatrix} \underline{d}_{t} \\ \overline{d}_{t} \end{bmatrix}
\end{align*}
with $\underline{C},\overline{C},\underline{d},\overline{d}$ taken from CROWN as \eqref{eq:crown} on $(\underline{x}_t,\overline{x}_t)$, and $\underline{M} = A + B^+\underline{C} + B^-\overline{C}$ and $\overline{M} = A + B^+\overline{C} + B^-\underline{C}$.

We use the neural network from \cite{ME-GH-CS-JPH:21} ($2\times 10\times 5\times 1$, ReLU). The performance of Algorithm \ref{alg:main} for various choices of $\varepsilon$, $D_p$, and $D_\text{N}$ is shown in Figure \ref{fig:DI_fig1}, for the initial set $[2.5,3]\times[-0.25,0.25]$ and a final time of $T=5$.

\paragraph*{Comparing with the literature}
Additionally, we compare the proposed ReachMM-CG to state-of-the-art partitioning algorithms for linear discrete-time systems: ReachMM \cite{SJ-AH-SC:23} (uniform initial partitioning), ReachLP-Uniform \cite{ME-GH-CS-JPH:21} (ReachLP with uniform initial partitioning), ReachLP-GSG (ReachLP with greedy sim-guided partitioning) \cite{ME-GH-CS-JPH:21}, and ReachLipBnB \cite{TE-SS-MF:22} (branch-and-bound using LipSDP \cite{MF-MM-GJP:22}). \

\begin{figure}
    \centering
    \includegraphics[width=0.95\linewidth]{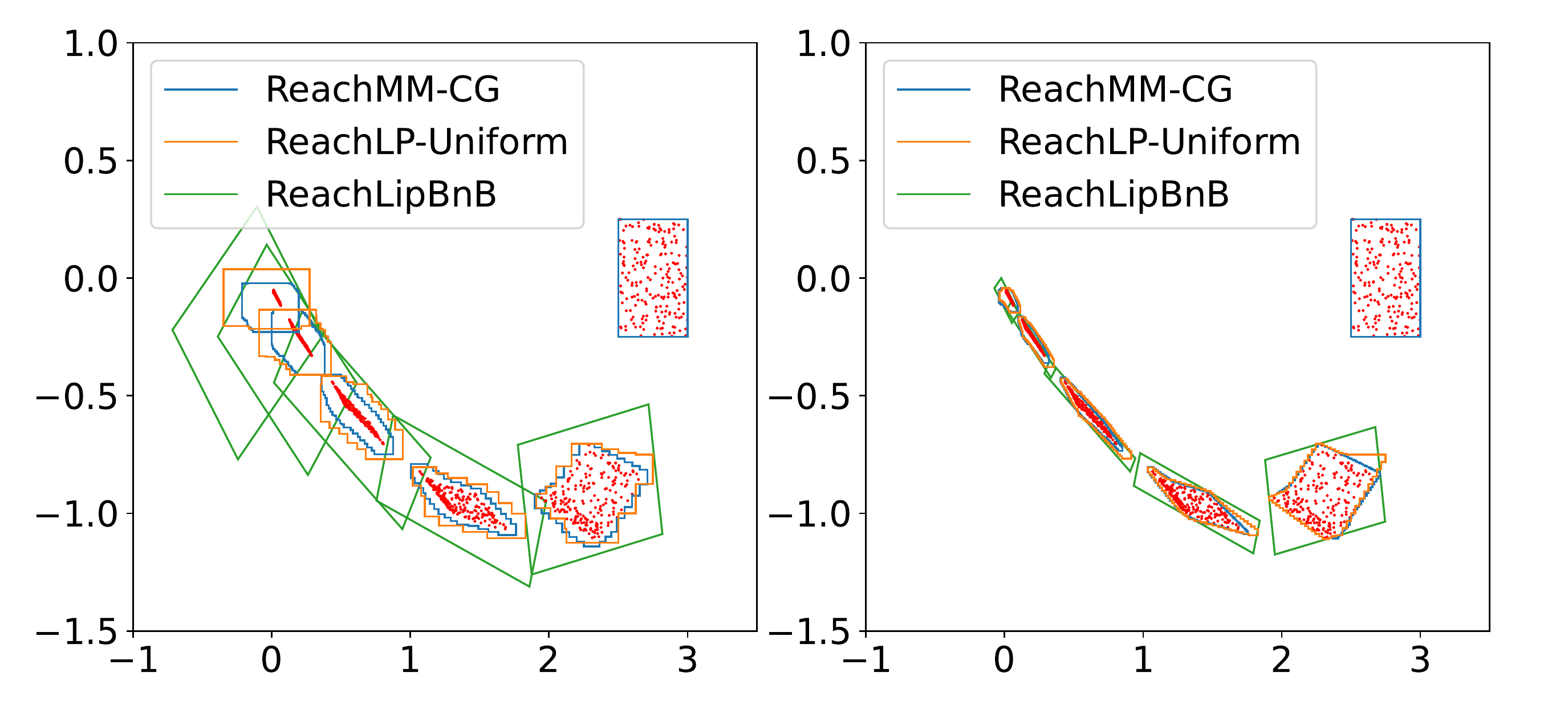}
    \caption{The over-approximated reachable sets of the closed-loop double integrator model~\eqref{eq:doubleintegrator} are compared for three different algorithms on two different runtime regimes, for the initial set $[2.5,3]\times[-0.25,0.25]$ and final time $T=5$. The experiment setup and performances are reported in Table~\ref{tab:DI_table}. 200 true trajectories are shown in {\color{tab:red} red}. The horizontal axis is $x_1$ and the vertical axis is $x_2$.}
    \label{fig:DI_fig3}
    \vspace{-1em}
\end{figure}

Each algorithm is run with two different sets of hyper-parameters, aiming to compare their performances across various regimes. The setup for ReachMM-CG is $(\varepsilon,\,D_p,\,D_\text{N})$; ReachMM is $(D_p,\,D_{\text{N}})$; ReachLP-Uniform is $\#$ initial partitions, ReachLP-GSG is $\#$ of total propogator calls, ReachLipBnB is $\varepsilon$. Their runtimes are averaged over 100 runs, with mean and standard deviation reported. The true areas of the reachable sets are computed using Python packages (Shapely, polytope).  The performance is outlined in Table~\ref{tab:DI_table}, and notable reachable sets are displayed in Figure~\ref{fig:DI_fig3}. ReachMM-CG outperforms SOTA across the board: for both setups, ReachMM-CG is significantly faster than the other methods while returning a tighter reachable set.

\begin{table}[h!]
    \centering
    \begin{tabular}{||c c c c||}
        \hline
        Method & Setup & Runtime (s) & Area \\
        \hline\hline
{\footnotesize ReachMM-CG} & $(0.1$, $3$, $1)$  & $\mathbf{0.079\pm0.001}$ & $\mathbf{1.0\cdot10^{-1}}$ \\
{\footnotesize (our method)}& $(0.05$, $6$, $2)$ & $\mathbf{0.833\pm0.025}$ & $\mathbf{7.5\cdot10^{-3}}$ \\
        \hline
        \multirow{2}{*}{\footnotesize ReachMM} 
& $(2$, $2)$    & $0.259\pm0.001$ & $1.5\cdot10^{-1}$ \\
& $(6$, $2)$    & $1.466\pm0.022$ & $9.0\cdot10^{-3}$ \\
        \hline
        \multirow{2}{*}{\footnotesize ReachLP-Unif} 
        & 4 & $0.212 \pm 0.002$ & $1.5\cdot10^{-1}$ \\
        & 16 & $3.149 \pm 0.004$ & $1.0\cdot10^{-2}$ \\
        \hline
        \multirow{2}{*}{\footnotesize ReachLP-GSG} 
        &  55 & $0.913 \pm 0.031$ & $5.3\cdot10^{-1}$\\
        & 205 & $2.164 \pm 0.042$ & $8.8\cdot10^{-2}$\\
        \hline
        \multirow{2}{*}{\footnotesize ReachLipBnB} 
        & $0.1$ & $0.956 \pm 0.067$ & $5.4\cdot10^{-1}$ \\
        & $0.001$ & $3.681 \pm 0.100$ & $1.2\cdot10^{-2}$ \\
        \hline
    \end{tabular}
    \caption{The performance of ReachMM-CG on the LTI Discrete-Time double integrator model, compared to other state of the art partitioning algorithms.}
    \label{tab:DI_table}
    \vspace{-1.5em}
\end{table}

\section{Conclusions}
In this paper, we propose an adaptive partitioning approach for interval reachability analysis of neural network controlled systems. The algorithm uses \textit{separation} of the neural network verifier and the dynamical system reachability tool, is \textit{spatially aware} in choosing the right locations of the state space to partition, and is \textit{temporally aware} to partition along trajectories rather than merely on the initial set. Using contraction theory for mixed monotone reachability analysis, we provide formal guarantees for the algorithm's performance. Finally, using simulations, we demonstrate significant improvement in runtime, accuracy, and generality as compared to existing partitioning approaches.

\bibliographystyle{ieeetr}
\bibliography{references,SJ}

\end{document}